\documentclass[12pt]{article}

\usepackage{graphicx}
\begin{document}

\begin{center}
{\bf Nonlinear $\arcsin$-electrodynamics and asymptotic Reissner-Nordstr\"om black holes} \\
\vspace{5mm} S. I. Kruglov
\footnote{E-mail: serguei.krouglov@utoronto.ca}
\underline{}
\vspace{3mm}

\textit{Department of Chemical and Physical Sciences, University of Toronto,\\
3359 Mississauga Road North, Mississauga, Ontario L5L 1C6, Canada} \\
\vspace{5mm}
\end{center}

\begin{abstract}
A model of nonlinear electrodynamics with the Lagrangian density ${\cal L} = -(1/\beta)\arcsin(\beta
F_{\mu\nu}F^{\mu\nu}/4)$ is proposed. The scale invariance and the dual invariance of electromagnetic fields are broken
in the model. In the limit $\beta\rightarrow 0$ one comes to Maxwell's electrodynamics and the scale and dual
invariances are recovered. We investigate the effect of coupling electromagnetic fields with the gravitational field.
The asymptotic black hole solution is found which is similar to the Reissner-Nordstr\"om solution.
We obtain corrections to Coulomb's law and to the Reissner-Nordstr\"om solution in the model proposed. The existence of the regular asymptotic at $r\rightarrow 0$ was demonstrated. The mass of the black hole is calculated possessing the electromagnetic origin.
It was shown that there are not superluminal fluctuations and principles of causality and unitarity take place.
\end{abstract}

\section{Introduction}

In cosmology the problems of the initial big bang singularity and inflation can be solved by means of nonlinear electrodynamics (NED). NED is described by classical theory and it may be investigated in the framework of the theory of gravity. Electromagnetic fields and gravitational fields in the early time of the universe creation were very strong, and nonlinear effects played very important role. NED for weak fields leads to Maxwell electrodynamics that can be considered as an approximation. Classical electrodynamics should be modified for strong electromagnetic fields \cite{Jackson} when the self-interaction of photons is important. In NED proposed by Born and Infeld \cite{Born} the total electric energy of charged particles is finite because of an upper limit on the electric field, and there is no singularity of the electric field at the origin of charged particles. It should be mentioned that Born-Infeld (BI) action arises from low energy effective action of the open superstring theory \cite{Fradkin}, \cite{Tseytlin}. It was also shown that Maxwell and BI theories bear both dual invariance \cite{Gibbons}, \cite{Gibbons1} (see also \cite{Garcia1}). Some models of NED were proposed where there is no singularity of the electric field at the origin of charged particles and the total electromagnetic energy is finite \cite{Shabad}, \cite{Kruglov0}, \cite{Kruglov}, \cite{Kruglov2}.
The finiteness of the electric energy of charged particles makes it passible to treat the mass of the electron as a pure electromagnetic energy. Thus, Dirac proposed a model of the electron \cite{Dirac} possessing a finite size and introducing an electromagnetic field source as well a source of a non-electromagnetic origin.

The regular exact black hole solution in General Relativity (GR) was represented in \cite{Ayon} where the source of gravity is a nonlinear electrodynamic field. Authors described
the system using not the Lagrangian but another function, obtained by means of a Legendre transformation \cite{Garcia1}. Other models of NED coupled to gravity possessing regular solutions were investigated in  \cite{Breton}, \cite{Dymnikova1}, \cite{Balard}, \cite{Kruglov5}. These  models have a singularity-free black hole solutions which behave asymptotically as the Reissner–Nordstr\"{o}m solution. It should be mentioned that any Lagrangian of NED coupled to gravity for regular electrically charged solutions is unavoidable branching \cite{Bronnikov}.

In this paper we consider a new NED model in the framework of the Lagrange dynamics, depending on a dimensional constant $\beta$, where electromagnetic fields are coupled to the gravitational field. At the limit $\beta\rightarrow 0$ the black hole geometry approaches to Einstein-Maxwell geometry. We study new Einstein-NED solution which gives corrections to the Coulomb law and to the Reissner-Nordstr\"om (RN) black hole solution and modifies the RN geometry.
The solution obtained gives new asymptotic terms at $r\rightarrow \infty$ for the Coulomb law and for the RN black hole metric in the order of $r^{-10}$.

In GR NED is of interest as processes of vacuum polarizations should be taken into consideration due to strong electromagnetic fields in the early epoch of the universe, at the origin of charged particles, and at the center of charged black holes. Instead of introducing the dark energy one can use nonlinear electromagnetic fields that influence on the evolution of the early universe near the Planck era and lead to inflation \cite{Garcia}, \cite{Camara}.
Some models of NED were considered in \cite{Elizalde}-\cite{Kruglov6} to describe accelerated expansion of the universe. In such models the magnetic universe is considered and magnetic fields undergo universe inflation. The stochastic magnetic fields are present and lead to the condition $(\rho + 3p)<0$ at the early universe epoch that gives the acceleration due to the Friedmann equation.

The cosmological constant $\Lambda$, in the $\Lambda$-Cold Dark Matter ($\Lambda$CDM) model, drives the present cosmic acceleration. At the same time the nonzero trace of the energy-momentum tensor in NED can mimic the cosmological constant \cite{Labun}, \cite{Schutzhold} at the early epoch because at the early universe NED describes strong self-interacting fields. At the present time electromagnetic fields are strong at the origin of charged particles and in the central regions of charged black holes.
For weak electromagnetic fields NED is converted into Maxwell electrodynamics possessing zero energy-momentum trace. The Einstein-Born-Infeld equations that take into consideration nonlinear effects were studied in \cite{Hendi}. One-loop quantum corrections to Maxwell electrodynamics lead to nonlinear Heisenberg-Euler Lagrangian \cite{Heisenberg}, \cite{Schwinger}, \cite{Adler}. In \cite{Hiroki} the static and spherically symmetric spacetime of black holes with the Heisenberg-Euler NED coupled to gravity was studied.

The paper is organized as follows. In Sec. 2 we consider a model of NED with the dimensional parameter $\beta$ and
obtain the energy-momentum tensor. The electric permittivity, $\varepsilon$, and the magnetic permeability, $\mu$,
depending on the electromagnetic fields are found. We show that the scale invariance and the dual invariance are broken
in the model. It is verified that WEC, DEC, and SEC are satisfied. The general principles of causality and unitarity were investigated. In Sec. 3 nonlinear electromagnetic fields coupled to gravity are studied. The corrections to the Coulomb law are found. The P framework was considered corresponding to electric-magnetic duality transformations.
It is shown that the electric field possesses the maximum at the origin of charged particles. In Sec. 4 the mass of the black hole is calculated possessing the electromagnetic origin. We obtain the black hole solution having the asymptotic Reissner-Nordstr\"om solution and find corrections to the Reissner-Nordstr\"om solution. The existence of the regular asymptotic of the metric function at $r\rightarrow 0$ was shown. A sound speed was calculated and it was shown that there are not superluminal fluctuations. A bound on an electric field was obtained to guarantee a classical stability. In Sec. 5 we make a conclusion.

We use the units with $c=\hbar=1$ and the metric $\eta=\mbox{diag}(-1,1,1,1)$.

\section{Nonlinear $\arcsin$-electrodynamics}

We propose NED with the Lagrangian density
\begin{equation}
{\cal L} = -\frac{1}{\beta}\arcsin(\beta{\cal F}),
 \label{1}
\end{equation}
where $\beta$ ($\beta>0$) is dimensional parameter with the dimension of (length)$^4$ and $\beta{\cal F}$ is dimensionless,
$F_{\mu\nu}=\partial_\mu A_\nu-\partial_\nu A_\mu$ is the field strength and
${\cal F}=(1/4)F_{\mu\nu}F^{\mu\nu}=(\textbf{B}^2-\textbf{E}^2)/2$. The model (1) is the modification of the model with two parameters proposed in \cite{Kruglov2}.
It will be shown that the parameter $\beta$ defines the maximum of the electric field at the origin of charged particles. For weak electromagnetic fields $\beta{\cal F}\ll 1$, using the Taylor series, we obtain from Eq. (1) the expression
\[
{\cal L}= - {\cal F}-\frac{\beta^2}{6}{\cal F}^3+{\cal O}(\beta^4{\cal F}^5).
\]
At $\beta{\cal F}\rightarrow 0$ (or $\beta\rightarrow 0$) the Lagrangian density (1) approaches to the Maxwell
Lagrangian density, ${\cal L}\rightarrow - {\cal F}$.
We find the symmetric energy-momentum tensor by varying the action corresponding to the Lagrangian density (1) with
respect to the metric tensor $g^{\mu\nu}$ \cite{Birula}
\begin{equation}
T^{\mu\nu}=H^{\mu\lambda}F^\nu_{~\lambda}-g^{\mu\nu}{\cal L},
\label{2}
\end{equation}
where
\begin{equation}
H^{\mu\lambda}=\frac{\partial {\cal L}}{\partial F_{\mu\lambda}}=\frac{\partial {\cal L}}{{\partial\cal
F}}F^{\mu\lambda}=-\frac{F^{\mu\lambda}}{\sqrt{1-\left(\beta {\cal F}\right)^2}}.
\label{3}
\end{equation}
The symmetric energy-momentum tensor obtained from Eqs. (2),(3) is given by
\begin{equation}
T^{\mu\nu}=-\frac{F^{\mu\lambda}F^\nu_{~\lambda}}{\sqrt{1-\left(\beta {\cal F}\right)^2}}-g^{\mu\nu}{\cal L}.
\label{4}
\end{equation}
One can find the trace of the energy-momentum tensor (4),
\begin{equation}
{\cal T}\equiv T_{\mu}^{~\mu}=\frac{4}{\beta}\arcsin(\beta{\cal F})- \frac{4{\cal F}}{\sqrt{1-\left(\beta {\cal
F}\right)^2}}.
\label{5}
\end{equation}
If $\beta\rightarrow 0$ we come to Maxwell's electrodynamics, and trace (5) becomes zero, ${\cal T}\rightarrow 0$.
The
scale invariance is broken because of nonzero trace of the energy-momentum tensor. The dilatation current is given
by
the expression $D_\mu=x_\nu T_{\mu}^{~\nu}$ and the divergence becomes $\partial_\mu D^\mu={\cal T}$.
The electric displacement field can be obtained from the expression $\textbf{D}=\partial{\cal L}/\partial
\textbf{E}$.
Then from Eq. (1) one finds the electric displacement field
\begin{equation}
\textbf{D}=\frac{\textbf{E}}{\sqrt{1-\left(\beta {\cal F}\right)^2}}.
\label{6}
\end{equation}
With the help of the relation $\textbf{D}=\varepsilon \textbf{E}$, we obtain the electric permittivity
\begin{equation}
\varepsilon=\frac{1}{\sqrt{1-\left(\beta {\cal F}\right)^2}}.
\label{7}
\end{equation}
From the definition $\textbf{H}=-\partial{\cal L}/\partial \textbf{B}$ one finds the magnetic field
\begin{equation}
\textbf{H}= \frac{\textbf{B}}{\sqrt{1-\left(\beta {\cal F}\right)^2}}.
\label{8}
\end{equation}
From the equality $\textbf{B}=\mu \textbf{H}$ the magnetic permeability is equal to $\mu=1/\varepsilon$. From Eqs.
(6),(8) we observe that the relation $\textbf{D}\cdot\textbf{H}=\varepsilon^2\textbf{E}\cdot\textbf{B}$ holds. As a
result, $\textbf{D}\cdot\textbf{H}\neq\textbf{E}\cdot\textbf{B}$, and according to \cite{Gibbons} (see also \cite{Plebanski}) we make a conclusion that the dual symmetry is violated in our model.
By virtue of Eqs. (6),(8) the Lagrange-Euler equations can be represented in the form of the first pair of the
Maxwell equations
\begin{equation}
\nabla\cdot \textbf{D}= 0,~~~~ \frac{\partial\textbf{D}}{\partial
t}-\nabla\times\textbf{H}=0.
\label{9}
\end{equation}
The equation $\partial_\mu \widetilde{F}_{\mu\nu}=0$, where $\widetilde{F}_{\mu\nu}$ is a dual tensor, follows from the Bianchi identity and leads to the second pair of Maxwell's equations
\begin{equation}
\nabla\cdot \textbf{B}= 0,~~~~ \frac{\partial\textbf{B}}{\partial
t}+\nabla\times\textbf{E}=0.
\label{10}
\end{equation}
Eqs. (6), (8), (9), (10) represent the nonlinear Maxwell equations because the electric permittivity $\varepsilon$
and the magnetic permeability $\mu$ depend on the electromagnetic fields $\textbf{E}$, $\textbf{B}$.
We mention that only specific functions ${\cal L({\cal F})}$ lead to general principles of causality and unitarity. These principles guarantee that neither ghost, nor tachyons will appear. The causality principle is a requirement that the group velocity of elementary excitations over a background is less than the speed of light. It gives the restriction on the function ${\cal L}({\cal F})$: ${\cal L_{\cal F}} \leq 0$  (${\cal L_{\cal F}}=\partial {\cal L}/\partial {\cal F})$ \cite{Shabad1}. One can verify that this condition is satisfied for our model (1). The unitarity principle guarantees the positive definiteness of the norm of every elementary excitation of the vacuum and leads the the requirement ${\cal L}_{\cal F}+2{\cal F}{\cal L}_{{\cal F}{\cal F}}\leq 0$ \cite{Shabad1}. From Eq. (1), for pure electric field ($\textbf{B}=0$) which will be considered, we find ${\cal L}_{\cal F}+2{\cal F}{\cal L}_{{\cal F}{\cal F}}=-(1+\beta^2E^4/4)/(1-\beta^2E^4/4)^{3/2} < 0$. Thus, the unitarity principle is also satisfied for the model under consideration.

\subsection{Energy conditions}

The Weak Energy Condition (WEC) \cite{Hawking} is fulfilled if and only if the relations
\begin{equation}
\rho\geq 0,~~~\rho+p_m\geq 0 ~~(m=1,~2,~3)
\label{11}
\end{equation}
hold, where $\rho=T^0_{~0}$ is the energy density, and $p_m=-T^m_{~m}$ (there is no summation in the index $m$) are principal pressures. Eqs. (11) require that any local observer measures the energy density to be nonnegative.
It should be mentioned that WEC (equations (11)) is equivalent to the condition $T_{\mu\nu}\xi^\mu\xi^\nu\geq 0$ for any timelike vector $\xi^\nu$ \cite{Hawking}.
From Eq. (4), for the case $\textbf{B}=0$, we obtain
\[
\rho=\frac{E^2}{\sqrt{1-(\beta E^2/2)^2}}-\frac{1}{\beta}\arcsin\left(\frac{\beta E^2}{2}\right),
\]
\begin{equation}
p_m=\frac{1}{\beta}\arcsin\left(\frac{\beta E^2}{2}\right)-\frac{E_m^2}{\sqrt{1-(\beta E^2/2)^2}} ~~~(m=1,~2,~3),
\label{12}
\end{equation}
$E^2=E_1^2+E_2^2+E_3^2$. The plot of the function $\rho\beta$  vs $E\sqrt{\beta/2}$ is given in Fig. 1.
\begin{figure}[h]
\includegraphics[height=3.0in,width=3.0in]{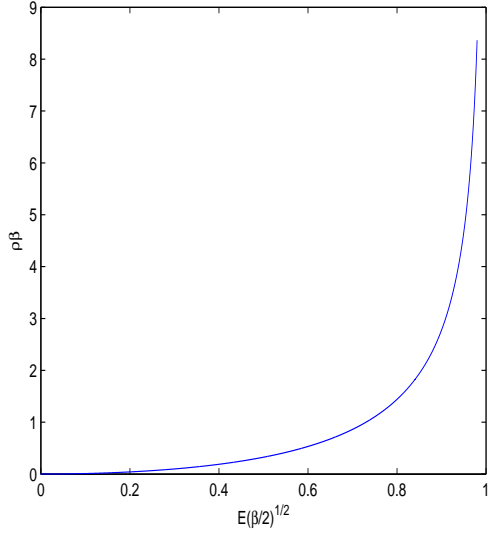}
\caption{\label{fig.1}The function  $\rho\beta$  vs $E\sqrt{\beta/2}$.}
\end{figure}
One can verify that WEC is satisfied for any values of the electric field (note that $\beta E^2/2<1$).

The Dominant Energy Condition (DEC) \cite{Hawking} is satisfied if and only if
\begin{equation}
\rho\geq 0,~~~\rho+p_m\geq 0,~~~\rho-p_m\geq 0~~(m=1,~2,~3).
\label{13}
\end{equation}
It is easy to verify, by virtue of Eqs. (12), that the inequalities (13) hold which guarantee that the speed of sound can not be greater than the speed of light.

The Strong Energy Condition (SEC) \cite{Hawking} holds if
\begin{equation}
\rho+\sum_{m=1}^3p_m\geq 0.
\label{14}
\end{equation}
One can check with the help of Eqs. (12) that SEC (14) is satisfied. The pressure for the Lagrangian density (1) can be obtained from the relation $p={\cal L}+(E^2/3){\cal L}_{\cal F}$. From Eqs. (1),(3) we find $p=(1/3)\sum_{m=1}^3p_m$. Then SEC becomes $\rho+3p\geq 0$ and tells that there can not be the acceleration of the universe in the model of NED coupled to GR.

\section{Nonlinear electromagnetic fields and black holes}

Let us consider NED described by the Lagrangian density (1) coupled with the gravitational field,
\begin{equation}
S=\int d^4x\sqrt{-g}\left[\frac{1}{2\kappa^2}R+ {\cal L}\right],
\label{15}
\end{equation}
where $R$ is the Ricci scalar, $\kappa=\sqrt{8\pi G}\equiv M_{Pl}^{-1}$, $G$ is the gravitation (Newton's) constant, and $M_{Pl}$ is the reduced Planck mass. One can explore units where $G=1$. From Eq. (15) we obtain the Einstein equation and equations for electromagnetic fields
\begin{equation}
R_{\mu\nu}-\frac{1}{2}g_{\mu\nu}R=-\kappa^2T_{\mu\nu},
\label{16}
\end{equation}
\begin{equation}
\partial_\mu\left(\frac{\sqrt{-g}F^{\mu\nu}}{\sqrt{1-\left(\beta {\cal F}\right)^2}}\right)=0.
\label{17}
\end{equation}
To find the static charged black hole solutions to Eqs. (16),(17) we use the spherically symmetric line element in
$(3+1)$-dimensional spacetime
\begin{equation}
ds^2=-f(r)dt^2+\frac{1}{f(r)}dr^2+r^2(d\vartheta^2+\sin^2\vartheta d\phi^2).
\label{18}
\end{equation}
We suppose that the vector-potential has non-zero component $A_0(r)$ so that ${\cal F}=-[E(r)]^2/2$ at $\textbf{B}=0$, and Eq. (17) is
rewritten as
\begin{equation}
\partial_r\left(\frac{2r^2 E(r)}{\sqrt{4-\beta^2 [E(r)]^4}}\right)=0.
\label{19}
\end{equation}
We obtain, after integrating Eq. (19), the equation
\begin{equation}
2r^2E(r)=Q\sqrt{4-\beta^2 [E(r)]^4},
\label{20}
\end{equation}
where $Q$ is the integration constant. One can introduce the dimensionless variables
\begin{equation}
y=\sqrt{\frac{\beta}{2}}E(r),~~~~x=\frac{r}{(\beta Q^2)^{1/4}}.
\label{21}
\end{equation}
We find the real solution to Eq. (20), with the notations of (21), as follows
\begin{equation}
y=\sqrt{\sqrt{x^8+1}-x^4}.
\label{22}
\end{equation}
The plot of the function $y(x)$ is represented by Fig. 2.
\begin{figure}[h]
\includegraphics[height=3.0in,width=3.0in]{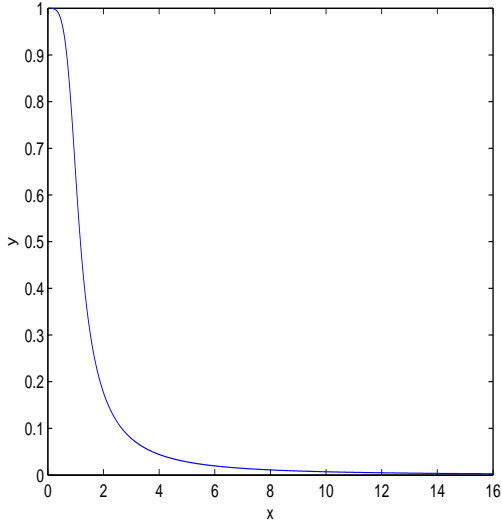}
\caption{\label{fig.2}The function  $y$  vs $x$.}
\end{figure}
One obtains the finite value $E(0)=\sqrt{2/\beta}$ ($y=1$) at the origin, and as a result, there is not singularity
of the electric field.
This is non-Maxwell behavior of the electric field in the center.
The Lagrangian density (1) possesses branching: the ${\cal L}({\cal F})$ is the multi-valued function.
But the Lagrangian density (1) as a function of $x$ (or $r$ according to Eq. (21)) is ${\cal L}=-(1/\beta)\arcsin(\sqrt{x^8+1}-x^4)$, represents the parametric form, and does not have a branching (see Fig. 3).
\begin{figure}[h]
\includegraphics[height=3.0in,width=3.0in]{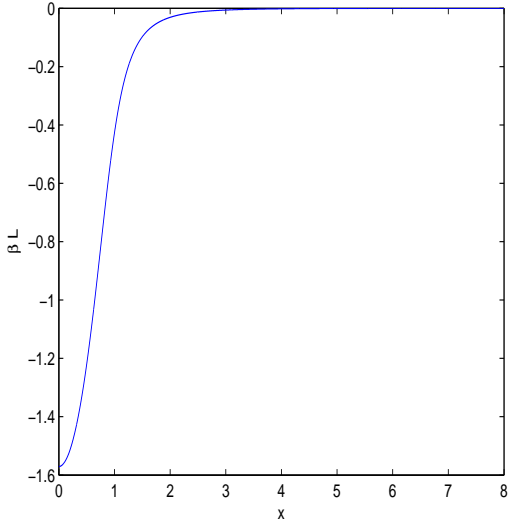}
\caption{\label{fig.3}The function  $\beta {\cal L}$  vs $x=r/(\beta Q^2)^{1/4}$.}
\end{figure}
In the following we investigate the dependence of the electric field and the electric potential as a function of $r$.
The Taylor series of the function $y(x)$ at $x\rightarrow\infty$ reads
\begin{equation}
y=\frac{1}{\sqrt{2}x^2}-\frac{1}{8\sqrt{2}x^{10}}+{\cal O}(x^{-13}).
\label{23}
\end{equation}
The asymptotic value of the electric field at $r\rightarrow \infty$ follows from Eqs. (21),(23)
\begin{equation}
E(r)=\frac{Q}{r^2}-\frac{Q^5\beta^2}{8r^{10}}+{\cal O}(r^{-13}).
\label{24}
\end{equation}
The first term in Eq. (24) corresponds to the Coulomb law with $Q$ being the charge. The second term in Eq. (24)
represents the  correction to the Coulomb law at $r\rightarrow\infty$. We see that at big distances the correction to
the Coulomb law is negligible. If $\beta=0$ we come to the Maxwell electrodynamics and the Coulomb law $E=Q/r^2$
holds. The electric potential, $A_0(r)=\int E(r)dr$, follows from Eqs. (21),(22)
\[
A_0(r)=\frac{\sqrt{2}}{Q\beta}\int dr\sqrt{\sqrt{r^8+Q^4\beta^2}-r^4}
\]
\[
=\frac{\sqrt{2}}{3Q\beta}\biggl[-2^{7/4}Q\sqrt{\beta}\sqrt[4]{\sqrt{r^8+Q^4\beta^2}-r^4}
\]
\[
\times F\left(\frac{1}{8},\frac{3}{4};
\frac{9}{8};\frac{(r^4-\sqrt{r^8+Q^4\beta^2})^2}{Q^4\beta^2}\right)
\]
\begin{equation}
+r\sqrt{\frac{r^8+Q^4\beta^2}{\sqrt{r^8+Q^4\beta^2}-r^4}}-\frac{r^5}{\sqrt{\sqrt{r^8+Q^4\beta^2}-r^4}}\biggr]+C,
\label{25}
\end{equation}
and $F(a,b;c;z)$ is the hypergeometric function (see \cite{Bateman}).
To simplify expression (25) we can use the approximate value of the potential $A_0(r)$ at $r\rightarrow\infty$. The
Taylor series of $A_0(r)$ at $r\rightarrow\infty$ (putting the integration constant to be zero, $C=0$) gives
\begin{equation}
A_0(r)=-\frac{Q}{r}+\frac{Q^5\beta^2}{72r^9}+{\cal O}(r^{-12}).
\label{26}
\end{equation}
From Eq. (26), after taking the derivative with respect to $r$, we obtain the expression for the electric field,
Eq. (24). Eq. (26) shows that corrections to the electric potential, due to nonlinear effects, are small at $r\rightarrow\infty$.
Assuming $r=0$ in Eq. (25), one finds the electric potential in the center
\begin{equation}
A_0(0)=-\frac{2^{9/4}\sqrt{Q}}{3\beta^{1/4}}F\left(\frac{1}{8},\frac{3}{4};\frac{9}{8};1\right)
=-\frac{\sqrt{Q}}{2^{3/4}3\beta^{1/4}}B\left(\frac{1}{8},\frac{1}{4}\right),
\label{27}
\end{equation}
where $B(x,y)=\Gamma(x)\Gamma(y)/\Gamma(x+y)$ is the beta-function \cite{Bateman}. One can use the approximate value
$B(1/8,1/4)\approx 11.523252$. Then the electric potential (27) becomes
\begin{equation}
A_0(0)\approx-\frac{2.28\sqrt{Q}}{\beta^{1/4}}.
\label{28}
\end{equation}
Thus, the electric potential, as well as the electric field, is finite at the origin of the charged object. As a
result, there are not singularities of the electric field and the electric potential at the origin of charged
objects. The cut-off of the electric potential (28) is due to non-Maxwell behavior of the original Lagrangian density (1) in the center $r=0$.
The similar situation occurs in well-known BI electrodynamics.

It should be noted that for the system obeying the spherical symmetry the radial electric field is $E(r)=E_1=F_{01}=-F_{10}$ and the radial magnetic field is $B(r)=F_{23}=-F_{32}$ \cite{Bronnikov}. Then the components of the energy-momentum tensor are $\rho=T^0_{~0}=T^1_{~1}=-p_r$ ($T^1_{~1}\equiv T^r_{~r}$). This leads to the condition $\rho+p_r=0$. The tangential pressure for perfect fluid is defined as $p_\perp =-T^\vartheta_{~\vartheta} =-T^\phi_{~\phi}$ \cite{Tolman}. It was shown from the conservation of the energy-momentum tensor \cite{Dymnikova1} that $p_\perp =-\rho-r\rho'/2$ ($\rho'=d\rho/dr$). We obtain from Eq. (12)
\[
\rho'(r)=\frac{2E'(r)E(r)[1+(\beta [E(r)]^2/2)^2]}{[1-(\beta [E(r)]^2/2)^2]^{3/2}}.
\]
The sign of $\rho'(r)$ depends on the sign of $E'(r)$. It follows from Eqs. (21),(22) and from Fig. 2 that $E'(r)\leq 0$, Therefore
$\rho'(r)\leq 0$ and the relation $p_\perp +\rho \geq 0$ holds. As a result, (for the case $\textbf{B}=0$) WEC as well as DEC and SEC are satisfied.

\subsection{P framework}

There is an alternative form of NED, gained by the Lagrangian using a Legendre transformation \cite{Garcia1}. Introducing the tensor $P_{\mu\nu}={\cal L}_{\cal F}F_{\mu\nu}/2$ in our model with the invariant
\begin{equation}
P=P_{\mu\nu}P^{\mu\nu}=\frac{{\cal F}}{1-(\beta {\cal F})^2},
\label{29}
\end{equation}
we consider the Hamilton-like variable
\begin{equation}
{\cal H}=2{\cal F}{\cal L}_{\cal F}-{\cal L}=\frac{1}{\beta}\arcsin\left(\beta {\cal F}\right)-\frac{2{\cal F}}{\sqrt{1-(\beta {\cal F})^2}}.
\label{30}
\end{equation}
One can verify that the quantity ${\cal H}$ coincides with the energy density (12), ${\cal H}=\rho$ and the relations \begin{equation}
{\cal L}_{\cal F}{\cal H}_P=1,~~~~P{\cal H}_P^2={\cal F},~~~~{\cal L}=2P{\cal H}_P-{\cal H}
\label{31}
\end{equation}
hold, where
\begin{equation}
{\cal H}_P=\frac{\partial {\cal H}}{\partial P}=-\sqrt{1-(\beta {\cal F})^2}.
\label{32}
\end{equation}
It follows from Eq. (29) that the function ${\cal F}(P)$ is monotonic function, and therefore, there is a one to one correspondence between two frames, ${\cal F}$ and $P$,  \cite{Bronnikov}. Thus, there is electric-magnetic duality between two frames. Any solution in ${\cal F}$-frame with the Lagrangian density ${\cal L}({\cal F})$ possesses a counterpart in the $P$-frame by the substitution ${\cal F}\rightarrow P$, ${\cal L}\rightarrow {\cal H}$, $F_{01}\rightarrow P_{23}$, $F_{23}\rightarrow P_{01}$, and conversely. This duality connects solutions of different theories with only the exception for Maxwell's theory, where ${\cal L}={\cal H}=-{\cal F}=-P$. Thus, the electric solution obtained in our model with the Lagrangian density (1) corresponds to the magnetic solution (with the duality transformation replacement) for the model with the function (30). One can see that for weak fields, $\beta{\cal F}\ll 1$, both models, (1) and (30) approach to the Maxwell theory, ${\cal L}=-{\cal F}$.

\section{Asymptotic Reissner-Nordstr\"{o}m black holes}

To find the Ricci scalar we use the relation that follows from Einstein's equation (16)
\begin{equation}
R=\kappa^2{\cal T},
\label{33}
\end{equation}
and the trace of the energy-momentum tensor is given by Eq. (5).
Then from Eqs. (5),(33) and the relation ${\cal F}=-(1/2)[E(r)]^2$, we obtain the Ricci scalar
\begin{equation}
R=\kappa^2\left[\frac{2[E(r)]^2}{\sqrt{1-\beta^2
[E(r)]^4/4}}-\frac{4}{\beta}\arcsin\left(\beta[E(r)]^2/2\right)\right].
\label{34}
\end{equation}
It follows from Eq. (34) that the Ricci scalar approaches to zero at $r\rightarrow \infty$, $\lim_{r\rightarrow \infty}R=0$, and spacetime becomes flat.
The plot of the function $R\beta/\kappa^2$  vs $r/(\beta Q^2)^{1/4}$ is given by Fig. 4.
\begin{figure}[h]
\includegraphics [height=3.0in,width=3.0in] {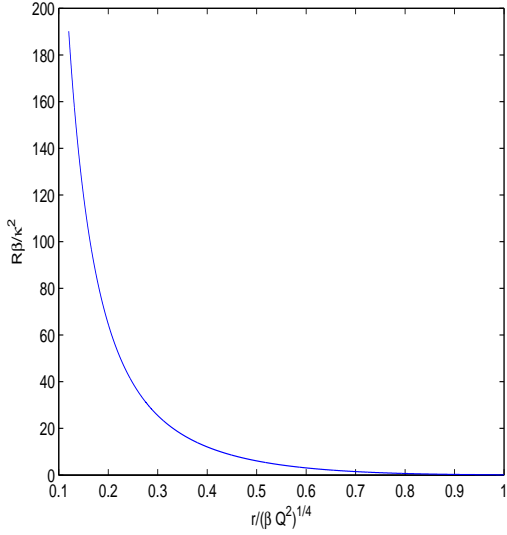}
\caption{\label{fig.4}The function  $R\beta/\kappa^2$  vs $r/(\beta Q^2)^{1/4}$.}
\end{figure}
One obtains the approximate value of $R$ using the Taylor series at $r\rightarrow\infty$ and taking $\sqrt{\beta} E(r)\ll 1$,
\begin{equation}
R=\kappa^2\left(\frac{\beta^2Q^6}{6r^{12}}+{\cal O}(r^{-13})\right).
\label{35}
\end{equation}
A static, spherically symmetric line element is given by Eq. (18),
where the metric function and mass function can be written in the form
\begin{equation}
f(r)=1-\frac{2GM(r)}{r},~~~~M(r)=\int_0^r\rho(r)r^2dr=m-\int^\infty_r\rho(r)r^2dr
\label{36}
\end{equation}
with the mass of the black hole $m$. By virtue of the energy density (12) one can represent the mass function as
\[
M(r)=m-Q^{3/2}\beta^{-1/4}\int_{r/(\beta Q^2)^{1/4}}^\infty \biggl[\sqrt{2\left(\sqrt{x^8+1}-x^4\right)}
\]
\begin{equation}
-x^2\arcsin\left(\sqrt{x^8+1}-x^4\right)\biggr]dx.
\label{37}
\end{equation}
We calculate the mass of the black hole
\begin{equation}
m=\int_0^\infty\rho(r)r^2dr\approx 1.52Q^{3/2}\beta^{-1/4}.
\label{38}
\end{equation}
The mass of the black hole has here the electromagnetic origin and is expressed through the constant $\beta$ introduced. The integral (37) possesses a complicated structure, and therefore, we use the asymptotic values.
At $r\rightarrow \infty$ ($x=r/( \beta Q^2)^{1/4}\rightarrow \infty$) one can get the asymptotic
\begin{equation}
\sqrt{2\left(\sqrt{x^8+1}-x^4\right)}-x^2\arcsin\left(\sqrt{x^8+1}-x^4\right)=\frac{1}{2x^2}
-\frac{1}{48x^{10}}+{\cal O}(x^{-13}).
\label{39}
\end{equation}
Replacing Eq. (39) into (37) and integrating, we obtain the asymptotic at $r\rightarrow\infty$ of the mass function and the metric
\begin{equation}
f(r)=1-\frac{2Gm}{r}+\frac{GQ^2}{r^2}-\frac{G\beta^2Q^6}{216r^{10}}+{\cal O}(r^{-13}).
\label{40}
\end{equation}
The last terms in Eq. (40) represent corrections to the Reissner-Nordstr\"{o}m solution.
Eq. (40) shows that spacetime asymptotically (at $r\rightarrow\infty$) is the Minkowski spacetime (flat). We have
Maxwell's electrodynamics at $\beta=0$ and then solution (40) is converted to the Reissner-Nordstr\"om solution. One
can see from Eq. (40) that corrections to the Reissner-Nordstr\"{o}m solution are very small at $r\rightarrow\infty$.
Other models of NED \cite{Breton}, \cite{Dymnikova1}, \cite{Hendi}, \cite{Kruglov5} also give asymptotic Reissner-Nordstr\"{o}m black hole solution but with different corrections. In our case corrections to the Reissner-Nordstr\"{o}m solution are negligible. We notice that corrections to the Reissner-Nordstr\"om solution lead to the change of the event horizon $r_+$ and the Cauchy horizon $r_-$ .
Regular electrically charged NED structures satisfying WEC, possess the de Sitter center \cite{Dymnikova1}. To get the asymptotic of the metric function at $r\rightarrow 0$ we consider the Taylor series of the expression at $r\rightarrow 0$
\begin{equation}
\sqrt{2\left(\sqrt{x^8+1}-x^4\right)}-x^2\arcsin\left(\sqrt{x^8+1}-x^4\right)=\sqrt{2}
-\frac{\pi x^2}{2}+{\cal O}(x^{4}).
\label{41}
\end{equation}
By virtue of Eqs. (36),(41) we obtain the metric function at $r\rightarrow 0$
\begin{equation}
f(r)=1-\frac{2\sqrt{2}GQ}{\sqrt{\beta}}+\frac{\pi G}{3\beta}r^2+{\cal O}(r^{4}).
\label{42}
\end{equation}
Eq. (42) shows the regular asymptotic of the metric function at $r\rightarrow 0$ with the negative cosmological constant $\Lambda=-\pi G/\beta$ corresponding to the anti-de Sitter space.

\subsection{Sound speed and causality}

The criteria that the speed of the sound is less that the light speed gives the restriction $c_s\leq 1$ \cite{Quiros}. The square sound speed has to be positive, $c^2_s> 0$, to meet the requirement of a classical stability. The square sound speed (at $B=0$) by taking into account Eq. (12) and the relation $p=(1/3)\sum_{m=1}^3p_m$, is given by
\begin{equation}
c^2_s=\frac{dp}{d\rho}=\frac{dp/d{\cal F}}{d\rho/d{\cal F}}=\frac{4-3\beta^2 E^4}{3(\beta^2 E^4+4)}.
\label{43}
\end{equation}
It is seen from Eq. (43) that $c_s^2\leq 1$ for any value of the electric field $E$, and therefore, there are not superluminal fluctuations. The requirement of a classical stability, $c^2_s> 0$, leads to the restriction
\begin{equation}
E\leq  \sqrt[4]{\frac{4}{3{\beta^2}}}.
\label{44}
\end{equation}
At the condition (44) the energy density perturbations do not uncontrollably grow. The bound value (44) is a little less than the maximum of the electric field, $E(0)=\sqrt{2/\beta}$, at the origin of charged particles. Thus, a bound (44) on an electric field  guarantees a classical stability.

\section{Conclusion}

We have investigated a new model of NED with the dimensional parameter $\beta$ and obtained the energy-momentum tensor and its nonzero trace. The scale invariance and the dual symmetry of the electromagnetic fields are violated because of the nonzero trace of the energy-momentum tensor. We show that WEC, DEC, and SEC are satisfied. The general principles of causality and unitarity were studied. Nonlinear electromagnetic fields coupled with the gravitational field have been studied. We have obtained the static
spherically symmetric solutions that correspond to the charged black holes. It was demonstrated that the electric
field has maximum at the origin of charged objects and there is no singularity. The electric potential is also finite
in the center. The P framework and the electric-magnetic duality transformations were considered.
We show that the black hole solution obtained possesses some corrections to the Reissner-Nordstr\"om
solution and asymptotically approaches to it. The mass of the black hole is evaluated possessing the electromagnetic origin. At $\beta \rightarrow 0$ NED becomes the Maxwell electrodynamics, and
we have the Reissner-Nordstr\"om solution. We show the existence of the nonsingular asymptotic of the metric function at $r\rightarrow 0$. We have calculated a speed of sound and it was shown that there are not superluminal fluctuations. A bound on an electric field was found that guarantees a classical stability.
The particular NED proposed has some attractive features:

i) the finiteness of the electric field at the origin,

ii) corrections to Coulomb's law and the Reissner-Nordstr\"om solution are very small (in the order of $r^{-10}$ at $r\rightarrow\infty$),

iii) the Minkowskian limit follows at $r\rightarrow\infty$ ($R\rightarrow 0)$,

iiii) the existence of the regular asymptotic at $r\rightarrow 0$.

Now we compare BI model and arcsin-electrodynamics proposed. In both models there are not singularities of the electric field at the origin of charged objects and the total electric energy is finite. In BI model the duality symmetry occurs but in arcsin-electrodynamics the duality symmetry is broken. It should be mentioned that in QED due to quantum corrections (the Heisenberg-Euler Lagrangian) the duality symmetry is also broken \cite{Kruglov4}.
There is no birefringence effect in BI model and the phase velocity does not depend on the orientation of the external magnetic field. In arcsin-electrodynamics the ${\cal L}({\cal F})$ is the odd function and corrections to Maxwell;s electrodynamics are in the order of $(\beta{\cal F})^3$. Because the corrections in the order of $(\beta{\cal F})^2$ are absent in our model, there is no the birefringence effect  \cite{Kruglov4}. BI fields coupled to GR can not accelerate the universe \cite{Novello1}. The same property takes place in our model as SEC ($\rho+3p\geq 0$) holds. In BI model there is a serious causality issues \cite{Quiros}. In the model under consideration only the stability problem arises when the electric field approaches to the value $E\rightarrow  \sqrt[4]{4/(3\beta^2)}$.
Because of some similarities and differences between BI model and arcsin-electrodynamics, the model proposed is of definite interest.

The investigation of Maxwell's electromagnetic fields coupled with gravitational fields non-linearly was done in
\cite{Odintsov}. One may study non-minimal coupling nonlinear electromagnetic fields considered with the gravitational
fields. This is of interest as inflation can be realized due to the non-minimal gravitational coupling of
the electromagnetic field, and that large-scale magnetic fields can be generated.
We also mention that there are instabilities of black holes in GR with the account of conformal anomaly \cite{Cai}. In our model the conformal symmetry is broken because of the violation of the scale invariance, and therefore, one may expect instabilities of black holes. The investigation of the instabilities of charged black holes and evaporation effects in our model can be studied (see the investigation of these effects in the Maxwell-$F(R)$ theory \cite{Odintsov1}).

\end{document}